\documentclass[twocolumn,           
               showpacs,            
               nopreprintnumbers,     
               aps,                 
               prd,          	    
               letterpaper,             
               groupeaddress,      
               nofootinbib,         
               tightenlines,        
               floats,floatfix      
               ]{revtex4-1}

\usepackage{graphicx}
\usepackage{dcolumn}
\usepackage{bm}
\usepackage{amsmath}
\usepackage{amsfonts,amssymb}
\usepackage{color}
\usepackage{enumerate}
\usepackage{float}
\DeclareMathOperator{\arcsinh}{arcsinh}

\begin{document}

\title{Cosmological constant problem in a scenario with compactifications (RS-I model)}

\author{C. Mart\'inez-Robles$^{1}$}
\email{cesar.martinez@fisica.uaz.edu.mx}

\author{Miguel A. Garc\'{\i}a-Aspeitia$^{1,2}$}
\email{aspeitia@fisica.uaz.edu.mx}

\affiliation{$^{1}$Unidad Acad\'emica de F\'isica, Universidad Aut\'onoma de Zacatecas, Calzada Solidaridad esquina con Paseo a la Bufa S/N C.P. 98060, Zacatecas, M\'exico}
\affiliation{$^{2}$Consejo Nacional de Ciencia y Tecnolog\'ia, \\ Av. Insurgentes Sur 1582. Colonia Cr\'edito Constructor, Del. Benito Ju\'arez C.P. 03940, M\'exico D.F. M\'exico}

\begin{abstract}
In this letter, we apply the Randall-Sundrum (RS) model, a scenario based on compactifications, to control the UV divergence of the zero-point energy density equation for the vacuum fluctuations, which has been unsuccessfully addressed to the cosmological constant (CC) due to a heavy discrepancy between theory and observation. Historically, the problem of CC has been shelved in the RS point of view, having few or non literature on the subject. In this sense and as done with the hierarchy problem, we apply the RS model to solve this difference via extra dimensions; we also describe how brane effects could be the solution to this substantial difference. It should be noticed that this problem is studied assuming first Minkoswki type  branes, and then followed by cosmologically more realistic FLRW type branes. We finally find some remarkably interesting consequences in the RS scenario: The CC problem can be solved via compactification of the extra dimension and the compactification radius turns out to be approximately twice the one used to solve the hierarchy problem in the $\beta\pi r$ factor by Randall and Sundrum, suggesting that this subtle difference in both problems could be caused by corrections that comes from quantum gravity effects. We also estimate the corresponding scale where, according to this results, we should begin to notice subtle deviations to the inverse square law of gravitation due to the presence of the extra dimension.
\end{abstract}

\keywords{Branes, Cosmological Constant, Dark Energy, Extra Dimensions}
\draft
\pacs{04.50.-h, 98.80.Jk}
\date{\today}
\maketitle


\emph{Introduction.---} In modern cosmology, dark energy (DE) is one of the most fundamental entities in the Universe, driving its recent accelerated evolution and being dominant compared to the other components of the Universe \cite{Ade:2015xua}. It is possible to notice that DE has the peculiar characteristic of accelerate the Universe, being one of the most fundamental discoveries of the last decade of XX century \cite{riess,*perlmutter}. The existence of DE is far from any doubt since its existence is imprinted in numerous cosmological observations like Cosmic Microwave Background Radiation (CMB) \cite{penziaswilson,*dicke}, Baryonic Acoustic Oscillations (BAO) \cite{eisenstein,*percival}, large scale structure formations \cite{spergelwmap,*spergelwmap2,*komatsuwmap,*hinshawwmap,*liddlestructure} among others \cite{stoughtonsloan,*abazajiansloan,*seljak,*tegmarksdsswmap}.

In this sense, the most simple solution to obtain an accelerated expansion comes from the addition of a positive cosmological constant (CC), but taking into account that there is a plethora of different approaches whose feasibility compete with the CC model. Some of the most important competitors are for example: Quintessence \cite{quintassence,*quintassence2}, Phantom Energy \cite{phantom}, Chaplygin gas \cite{chaplygingas,*chaplygingas2}, etc.., just to mention some of them. However other alternatives whose attempts to solve this problem comes from a more radical ideas of the space-time structure, mentioning some of them as: $f(R)$ gravities \cite{starobinsky}, DGP models \cite{dgp}, trans-Planckian models \cite{transplanck}, etc.., which are classified as modifications to gravity. Despite the efforts in this way, CC still remains as the best candidate to explain DE, supported by several cosmological observations (see for example \cite{hinshawwmap,*planck2014}), but maintaining the pathological problem of CC associated to the assumption of considering the quantum vacuum fluctuations as the source of its presence, since the result is $\sim118$ orders of magnitude higher than the value expected by observations. We should notice that this severe disagreement appears when we calculate the average energy density of vacuum fluctuations from the usual 4D theories  \cite{Carroll:2000fy,*zeldovich,*weinberg,*copeland}.

The classical recipe comes from the integral shown by Carroll or Copeland \emph{et. al.} \cite{Carroll:2000fy,copeland}, showing that the integral exhibit an ultraviolet divergence which grows as $\rho_{vac}\varpropto k^4$ in this limit, but can be corrected under the assumption that quantum field theory is valid up to some cut-off scale $k_{max}=m_{pl}$, where the sum is over all the zero-point energies of all normal modes of some field \cite{weinberg}. Under this consideration, we get 

\begin{equation}
\rho_{vac} = \frac{g}{4\pi^2} \int ^{k_{max}} _{0} dk k^{2} \sqrt{k^{2} + m^{2}} \approx10^{71}GeV^4, \label{eq1}
\end{equation}
where $g\equiv(-1)^{2j}(2j+1)$ is the number of degrees of freedom, $k_{max}\gg m$ for a field with spin $j>0$ and mass $m$, where $E(k)$ comes from the relativistic expression $\eta_{\mu\nu}k^{\mu}k^{\nu}=-m^2$, being $\eta_{\mu\nu}$ the Minkowski metric. In stark contrast, the expected results that comes from observations is $\rho_{obs}\approx10^{-47}GeV^4$ \cite{weinberg}.


In this vein, several ideas have been adopted aiming to finally unveil the solution to this elusive problem (see for example \cite{Collins,*Bailin,*Gibbons,*Maldacena,*Sorkin:2003bx,*Linde}), without any success so far. 
However, there exists an interesting extension to the General Theory of Relativity (GR), which is know as the Randall-Sundrum model (RSI-II) \cite{randallsundrum} with the capability of obtaining a natural solution to the hierarchy problem which afflicts the standard model of particles (SM). The solution is based on the idea of extending the number of dimensions to five, where the fifth dimension is know as the bulk, with a four-dimensional manifold embedded, known as brane. This idea can also be applied to the CC problem alleviating the quantum vacuum fluctuations through the anti-de Sitter bulk, remarking the subtle similarity between both problems and its possible relation with extra dimensions.

In addition to this, a recent paper by Hertzberg and Masoumi \cite{Hertzberg:2015bta}, discuss the usefulness of compactifications to solve the CC problem. Here the topological structure of vacuum is the key to solve the fine tuning problem. Notice that RS-type models fall into this category being an important candidate not only to solve the hierarchy problem but also the CC problem. 

From here, we will henceforth use natural units in which $\hbar=c=1$, unless otherwise stated.





\emph{Minkowski Branes Scenario.---}As we previously mentioned, the main achievement of RS model is solving the hierarchy problem through the following metric \cite{randallsundrum}:
\begin{equation}
ds^2=e^{-2\beta\vert y\vert}\eta_{\mu\nu}dx^{\mu}dx^{\nu}+dy^2,
\end{equation}
where $\beta=\sqrt{\tau^2/36M_{*}^6}$, $\tau$ is the brane tension and $M_{*}$ is the five-dimensional Planck mass. Under this consideration and the suppression of the actual vacuum as $v=e^{-\beta\pi r}\widehat{v}_{0}$, the physical components of the Higgs field appear at the TeV scale rather than at the Planck scale with $\beta r\sim12$ \cite{randallsundrum,PerezLorenzana:2004na}.

These ideas suggest that the same solution can be also applied to the CC as done with the hierarchy problem; however in the literature there is not a formal treatment to this problem and at most, some works only mention a possible solution, nevertheless, without clear foundations \cite{Chen:2007hy,*Das:2007qn,*Lahiri:2012tw}. 

In this regard, the proposed methodology is as follows: Under the assumption of RSI model and assuming that we are fixed in our brane (visible brane) \emph{i.e.}, $y=\pi r$, we have that the modified value of $E(k)$ is: $E^2=m^2e^{2\beta\pi r}+k^2$, where we have used $g_{\mu\nu}k^{\mu}k^{\nu}=-m^2$, being $g_{\mu\nu}$ the RS metric fixed on the brane. Then Eq. \eqref{eq1} can be rewritten as:
\begin{equation}
\rho_{vac} = \frac{g}{4\pi^2} \int ^{k_{max(RS)}} _{0} dk k^{2} \sqrt{k^{2} + m^2_{KK}}, \label{eq1RS}
\end{equation}
where for this case, the limits of the integral suppress the Planck mass as: $k_{max(RS)}=m_{pl}e^{-\beta\pi r}$ in similarity to the hierarchy problem for the Higgs field \cite{randallsundrum,PerezLorenzana:2004na}, being $m_{KK}=me^{\beta\pi r}$ the Kaluza-Klein (KK) mass. In addition, it is possible to notice that the ultraviolet divergence remains when it is evaluated in the infinity limit, then again it is necessary carry out an appropriate cut-off but now into the brane context. 
The first attempt is assuming $m_{KK} \ll k$, keeping in mind that the KK masses can play an important role in the result of the integral, which could invalidate this assumption. However, in a naive approximation and provisionally leaving aside such points, Eq. \eqref{eq1RS} can be straightforward integrated giving explicitly the following result:
\begin{equation}
\rho_{vac}=\frac{g}{16\pi^2} m_{pl}^4e^{-4\beta \pi r}. \label{firstint}
\end{equation}
In this case, the value of the free parameter must be $\beta r\sim23$, in order to solve the CC problem in this background and according to the reported bounds for the brane tension and the five-dimensional Planck mass, which are $\tau>1\rm MeV^4$ and $M_{*}>10^7\rm MeV$ respectively \cite{Maartens:2003tw}; this is almost twice the value reported for solving the hierarchy problem \cite{randallsundrum,PerezLorenzana:2004na}.

\begin{figure} [H]
\centering
\includegraphics[scale=0.5]{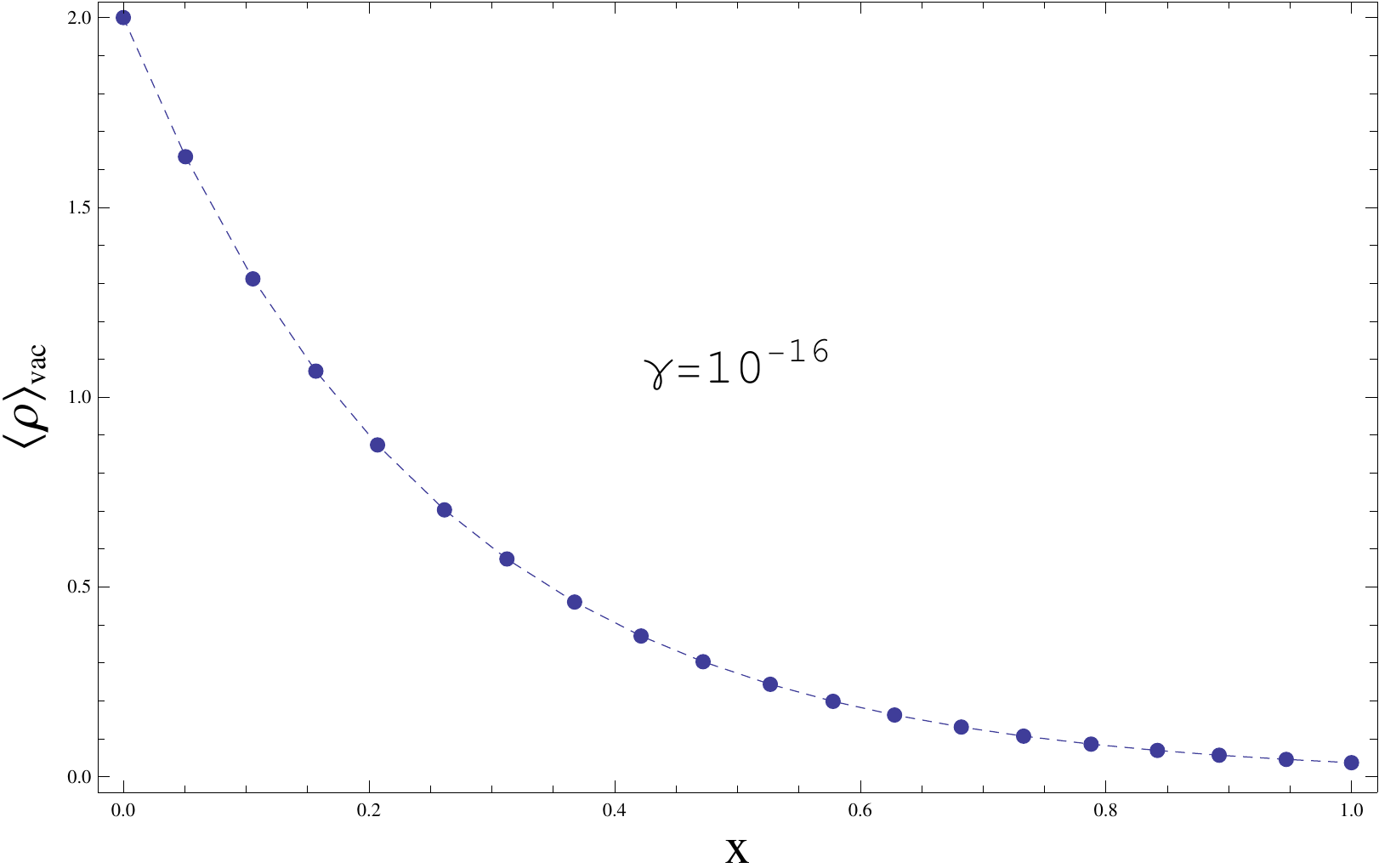} 
\caption{Monotonous behavior of normalized energy density for different values of the parameter $\rm x$, assuming $\gamma=10^{-16}$. It is important to notice that the values of $\rm x$ are always constant; showing only that for each value of $\rm x$, there is only \emph{one} associated value of the normalized energy density. See also how $\langle\rho\rangle_{vac}$ decrease for large values of the $\rm x$ parameter. See the text for more details.}
\label{FIG1}
\end{figure}

On the other hand, the full integral provided by Eq. \eqref{eq1RS} without extra considerations is:

\begin{eqnarray}
\langle\rho\rangle_{vac} &=& ( e^{-4\rm x}+\gamma^2)^{1/2}(2e^{-2\rm x}+\gamma^2e^{2\rm x}) \nonumber\\&-& \gamma^4e^{4\rm x}\arcsinh\left(\gamma^{-1}e^{-2\rm x}\right), \label{chida}
\end{eqnarray}
where $\langle\rho\rangle_{vac}\equiv32\pi^2\rho_{vac}/gm_{pl}^4$ is the dimensionless vacuum energy density, $\rm{x}\equiv$$\beta\pi r$ and $\gamma\equiv m/m_{pl}$. If the field mass is in the electroweak scale, \emph{i.e.} $m\sim1\rm TeV$ and $m_{pl}\sim10^{16}\rm TeV$ we have that $\gamma=10^{-16}$. The behavior of Eq. \eqref{chida} can be observed in Fig. \ref{FIG1}, for each point\footnote{Notice that Eq. \eqref{chida} is not a function, only we present the different values of the energy density as for different values of $\rm x$.}; where it must be noted that to get the value of $\rho_{vac}$, it is necessary to consider large values of $\beta r$.

If we also numerically solve Eq. \eqref{chida} for $\rm x$ when $\langle\rho\rangle_{vac} \sim 10^{-121}$, we obtain $\beta r \sim 23$; this solution suggests that our first approximation for the cut-off $m_{KK} \ll k$, is correct and we can use it to simplify calculations as we showed previously. This means that despite the differences in the relativistic energy equation found under the RS metric, we also find out that the momentum overcome the KK masses. From here, it is possible to establish the value for the compactification radius $r$, again, under the reported bounds for the brane tension and the five-dimensional Planck mass: $\tau>1\rm MeV^4$ and $M_{*}>10^7\rm MeV$ \cite{Maartens:2003tw} (see \cite{Germani:2001du,*Garcia-Aspeitia:2015mwa} for astrophysical bounds) then $r\gtrsim8.28\times10^{-4}\rm eV^{-1}\simeq0.16\rm nm$. Therefore, subtle deviations of the inverse square law of gravitation must be observed at these scales due to the presence of the extra dimension. However, experiments for deviations from Newton's law in 4 dimensions have't been tested at these scales yet \cite{Long:2002wn,*Kapner:2006si}.

To conclude this first part, it is necessary to discuss about the Lorentz invariance of the dispersion relation between the energy and momentum. The new dispersion relation is invariant under Lorentz transformations?

Notice that at second order approximation of the new dispersion relation we obtain:
\begin{equation}
E^2-k^2-m^2[1+2\beta\pi r+\mathcal{O}(r^2)]=0.
\end{equation}
From here we observe the addition of small perturbations to the traditional dispersion relation. In fact, it is possible to observe that the new terms does not depends of the energy and momentum in the form $E^2-k^2-m^2-\Pi(E,k)=0$ (see \cite{Collins:2004bp} for instance), which indeed suggests no violation of Lorentz invariance in this framework.


\emph{Cosmological Branes Scenario}.---As a complement, for RSI in a cosmological approach, it is possible to use a Gaussian normal coordinates in which the branes are fixed but the bulk metric is not manifestly static as \cite{Binetruy:1999hy}:
\begin{equation}
ds^2=-N^2(t,y)dt^2+A^2(t,y)(dx^i)^2+dy^2,
\end{equation}
where we take flat branes in consideration, being
\begin{subequations}
\begin{eqnarray}
N&=&\frac{\dot{A}(t,y)}{\dot{a}(t)}, \\
A&=&a(t)\left[\cosh \left(\beta y\right)-\left\lbrace1+\frac{\rho(t)}{\tau}\right\rbrace\sinh\left(\beta\vert y\vert\right)\right],
\end{eqnarray}
\end{subequations}
found by solving the 5D Einstein's equation \cite{Binetruy:1999hy}.

\begin{figure} [H]
\centering
\includegraphics[scale=0.4]{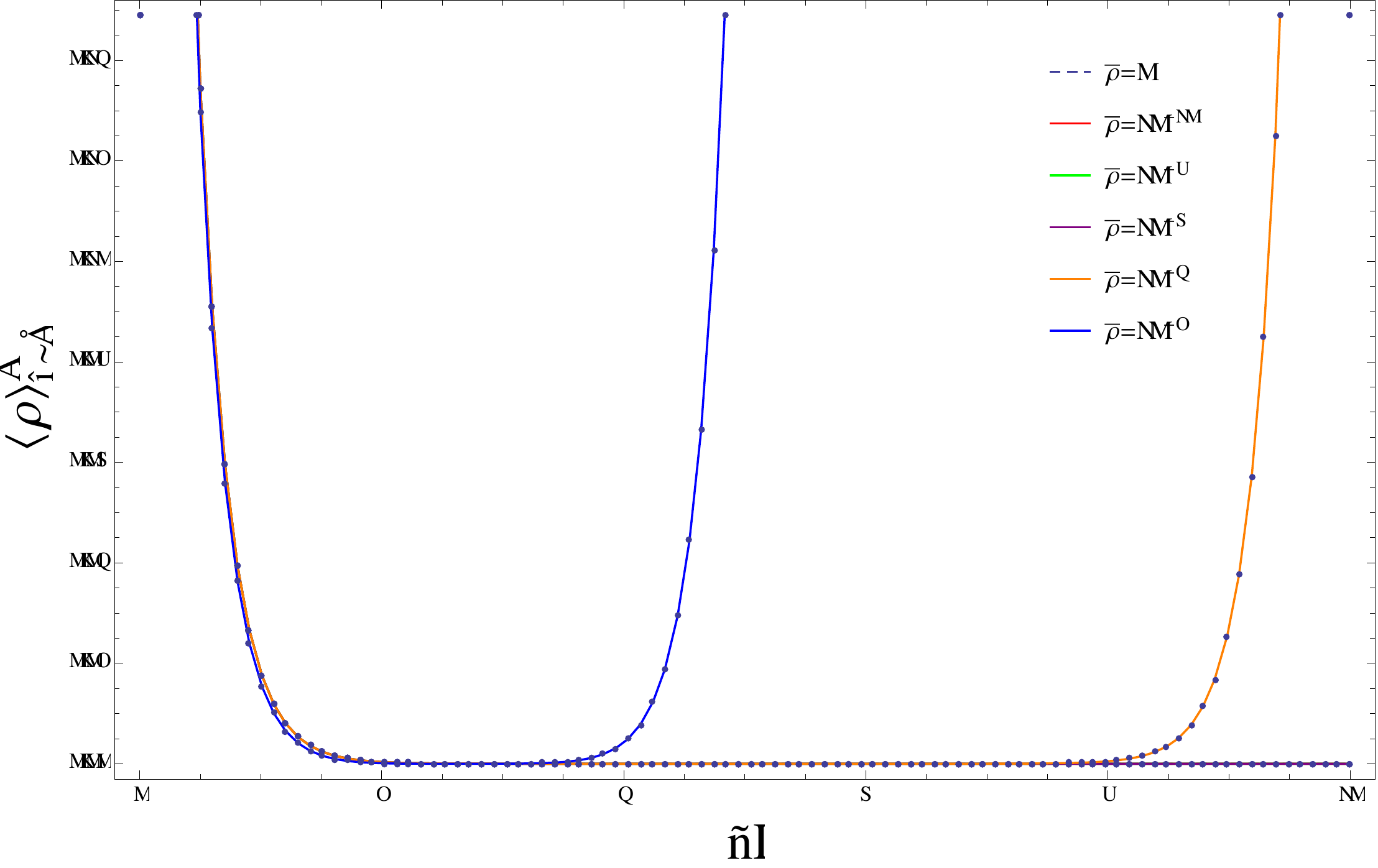} 
\caption{Behavior of Eq. \eqref{FIG2} for different values of $\bar{\rho}$; here it is always assumed that $\bar{\rho}\ll1$ which is expected from astrophysical and cosmological bounds. It is notorious how the presence of $\bar{\rho}$ affects the value of the energy density. It is also important to emphasize that there must be a minimum value of $\rm x^{\prime}$ where CC is recovered.}
\label{FIG2}
\end{figure}

The hidden brane it is fixed at $y=0$, recovering the FLRW metric; however for our brane, we have $y=\pi r$, obtaining a perturbed FLRW metric due to the small values of $\beta$. Then, it is possible to write the following dispersion relation for quantum vacuum in our brane as:
\begin{equation}
E=\mathcal{B}^{-1}\sqrt{a^2(t)\mathcal{B}^2k^2+m^2},
\end{equation}
which now depends on the scale factor which is time dependent and where 
\begin{eqnarray}
\mathcal{B}=\cosh \left(\beta \pi r\right)-\left[1+\frac{\rho_{crit}}{\tau}\right]\sinh\left(\beta\vert \pi r\vert\right),
\end{eqnarray}
is a function that rapidly decreases, and the ratio $\rho_{crit}/\tau$ must be small for recent times, where we fixed $\rho(t)=\rho(t_0)=\rho_{crit}$ as a constant, related with the critical density of our Universe. Now the integral for the quantum vacuum fluctuations can be written as:
\begin{equation}
\rho_{vac} = \frac{g}{4\pi^2\mathcal{B}} \int ^{k_C} _0 dk k^2 \sqrt{a^2\mathcal{B}^2k^2+m^2}, \label{CosmRS}
\end{equation}
proposing a new cut-off, $k_C=\mathcal{B}a^{-1/4}m_{Pl}$, with the aim of avoiding the time dependence caused by the scale factor. So now, the density energy for CC in our brane can be written as:
\begin{equation}
\langle\rho\rangle^{C}_{vac} = \left[\cosh \left(x^{\prime}\right)-(1+\bar{\rho})\sinh\left(\vert x^{\prime}\vert\right)\right]^4, \label{Cosmeq}
\end{equation}
where it is also considered that $a\mathcal{B}k\gg m$ and for $j>0$; similarly we define the following dimensionless variables $\langle\rho\rangle_{vac}^{C}\equiv16\pi^2\rho_{vac}/gm_{pl}^4$, where the super index $C$ refers to cosmology, $\bar{\rho}=\rho_{crit}/\tau$ and $\rm x$$^{\prime}\equiv\beta\pi r$, in order of avoid confusions with the Minkowski background. Thus, the behavior of Eq. \eqref{Cosmeq} can be observed in Fig. \ref{FIG2} for different values of $\bar{\rho}$. The reader should keep in mind that $\rm x^{\prime}$ does not vary; and $\langle\rho\rangle_{vac}$ only varies for different values of compactification radius. It is possible to observe that the value near to zero is maintained for small values of $\bar{\rho}$ which implies big values of the brane tension. Moreover we remark that in the cosmological case, large values of $\rm x^{\prime}$ and $\bar{\rho}$ present divergences in the value of the energy density which it is possible to observe in Fig. \ref{FIG2}; constraining the maximum value for $\rm x^{\prime}$ to solve the CC problem.

Constraints of brane tension through stellar and galactic dynamics together with cosmological bounds suggest $\tau>1\rm MeV^4$ \cite{Germani:2001du,Maartens:2003tw,Garcia-Aspeitia:2015mwa,Garcia-Aspeitia:2015eja,*Linares:2015fsa}, giving an extremely small value of the free parameter $\bar{\rho}=8.102h^2\times10^{-35}$, for recent values of critical density provided by observations \cite{Ade:2015xua}. Solving for $\rm x^{\prime}$ in Eq. \eqref{Cosmeq} we get $\beta r\sim22$ which is in extreme concordance with previous results; obtaining $r\gtrsim7.59\times10^{-4}eV^{-1}\simeq0.14\rm nm$ for the compactification radius.

\emph{Discussion}.---Cosmological constant is still the best candidate to explain DE behavior and its current Universe accelerated expansion, having more evidence of the necessity of DE as a fundamental component in our Universe and being CC the best solution. However as we previously stated, CC suffers of important problems whose solution must be pursued under new approaches. In this case RS models have shown their effectiveness solving fundamental problems like hierarchy in particle physics; despite the fact that its application to vacuum fluctuations associated with the energy density of CC, has been shelved in literature. 

In this sense we revisit this idea and formalize the mathematical fundaments to recover the expected value of the energy density of CC. As it happens with the Higgs field, the problem is reduced with the RS cut-off associated to the warp factor. Here we show the generalized results, controlling the quantum vacuum fluctuations with the free parameter $\beta r\sim23$ that  corresponds to a compactification radius of $r\simeq0.16\rm nm$. Therefore, deviations to Newton's law effects must be observed at these scales. Also it is important to mention that, Lorentz invariance is maintained in this context.

On the other hand, a cosmological background is studied obtaining the value of $\rm x^{\prime}$ to solve the CC problem and thus, the radion distance computed is $r\simeq0.14\rm nm$ which has a notable similarity with the results obtained for a Minkowski scenario.

There is a suggestive behavior with the corresponding values of the compactification radius $\beta r_{CC}$ and $\beta r_{Higgs}$ showing that for CC or quantum vacuum fluctuations it is almost twice the one obtained for  as the Higgs field, i.e. $r_{Higgs}\simeq0.8\rm nm$. However it is possible to observe that the difference is non significative being in the same order of magnitude. This suggest the existence of a unification framework, since both fundamental problems appear to have a common solution at the same scale. We also consider that the subtle differences between these two problems is due to quantum gravity effects which will finally be answered once we have a theory of gravitation at microscopic scales.

Other considerations like the radion stability should be taken into account, however it is remarkable that the two problems can be solved under the same approach with subtle differences between the two problems. Indeed, this approach can be extended to the problem of neutrino mass which is the same order of magnitude as the DE \cite{Simpson:2016gph} having a common origin. However this is still a speculation that should be studied in depth.

Moreover, future studies aim to analyze the most general expansion for a field at one-loop written as:
\begin{eqnarray}
\rho_{vac}&=&\frac{g}{16\pi^2}[c_1m_{pl}^4+c_2m_{pl}^2m^2+c_3m^4\ln\left(\frac{m^2}{m_{pl}^2}\right)\nonumber\\&+&...],
\end{eqnarray}
expanding in powers of $m/m_{pl}$, where $c_{1,2,3}=\mathcal{O}(1)$ are numbers that depend on the choice of regularization \cite{Hertzberg:2015bta}.

At the end, only recent observations will give us the definitive answer about the expected value for CC, responsible of the current acceleration of the Universe and perhaps future experiments of gravitation will determine the presence of extra dimensions compactified at small scales, However this approach is a suggestive point of view to definitely solve these fundamental problems.

\emph{Acknowledgements}. CM-R acknowledge support from CONACyT scholar fellowship. MAG-A acknowledge support from SNI-M\'exico and CONACyT research fellow. Instituto Avanzado de Cosmolog\'ia (IAC) collaborations.

\bibliography{absreferences}

\end{document}